# A Reasonable C++ Wrappered Java Native Interface

Craig Bordelon
*Telcordia Technologies, Piscataway, NJ 08854 USA*
*E-mail: cbordelon@acm.org*

**SUMMARY**

**A reasonable C++ Java Native Interface (JNI) technique termed C++ Wrappered JNI (C++WJ) is presented. The technique simplifies current error-prone JNI development by wrappering JNI calls. Provided development is done with the aid of a C++ compiler, C++WJ offers type checking and behind the scenes caching. A tool (jH) patterned on javah automates the creation of C++WJ classes.**

**The paper presents the rationale behind the choices that led to C++WJ. Handling of Java class and interface hierarchy including Java type downcasts is discussed. Efficiency considerations in the C++WJ lead to two flavors of C++ classes: jtypes and Jtypes. A jtype is a lightweight less than full wrapper of a JNI object reference. A Jtype is a heavyweight full wrapper of a JNI object reference.**

# Introduction

Java has emerged as an implementation language of choice for a variety of software applications. Still, at times there is a need to leverage pre-existing C/C++ software or to use low level system specific C/C++ software with Java software. Java and JDK 1.1/1.2 provide the Java `native` modifier and the Java Native Interface[6] (JNI) to allow C/C++ use with Java. The JNI, while powerful, lacks some things to make native C/C++ easy to write.

This paper rationalizes a technique to create C++ classes as needed to allow for simpler and at the same time efficient native C++ software. The created C++ classes encapsulate JNI calls in an `inline` manner and can sacrifice almost no performance. The created C++ classes allow a much better degree of compile time checking to help eliminate programmer errors.

The technique presented uses C++ to return to the simplicity of JDK 1.0 NMI (Native Method Interface) support. Only one member specific operation is needed to perform native side Java field access or method invoke. Yet, the advantages of JDK 1.1 JNI regarding portability remain. With JNI at the foundation, C++ is made to serve as an ideal language bridging native C (or C++) and Java. Think of C++ as the icing between Java and C/C++ layers of code.

Creating C++WJ classes is accomplished by the *jH* tool. This tool is briefly described as well as suggestions for its use. The standard *javah* tool if extended could provide options to accomplish the features of **jH**.

# JNI Refresher

JDK 1.1/1.2 JNI is a set of functions and type definitions for C and C++ code that allows interface to the Java Virtual Machine (JVM) with the means to:

- Access Java variables
- Invoke Java methods
- Create Java objects and optimize object lifetimes
- Synchronize code sections
- Throw and catch exceptions
- Start and end a JVM
- Interoperate with C/C++ data such as null terminated strings

JNI is abstracted from underlying JVM memory layouts and so compiled code with JNI function calls links and runs with any JVM for a particular hardware and compiler treatment (calling convention and C/C++ struct layouts). A function interface into necessary aspects of the JVM accomplishes the separation needed for this degree of portability. Figure 1 shows a Java program to create and print a `java.util.BitSet` with a single bit on whose bit position is the value of the first element of the input `java.lang.String[]` turned into an `int`. Figure 2 shows the equivalent C++ native code using JNI. A single *Java_* function provides the native main implementation. The `Integer.value` field although private is accessible with JNI. For simplicity, the code in Figure 2 disregards possible JNI function errors.

```
public class Bar {
    public static void main(String[] args){
        java.util.BitSet bs = new java.util.BitSet();
        bs.set(Integer.valueOf(args[0]).intValue());
        System.out.println(bs);
        }
    };
```

**Figure 1.** Example Java Program

```
public class Bar {
    public static native void main(String[] a);
    static {System.loadLibraray("foo");}
    };

//C++ from here
#include "Bar.h" //javah generated

JNIEXPORT void JNICALL Java_Bar_main(
    JNIEnv* e,jclass this_class,jobjectArray args){
    jclass jcB = e->FindClass("java/util/BitSet");
    jobject jbs = e->NewObject(jcB,e->GetMethodID(jcB,
        "<init>","()V"));
    jclass jcI = e->FindClass("java/lang/Integer");
    jobject jo = e->CallStaticObjectMethod(jcI,
        e->GetStaticMethodID(jcI,"valueOf",
            "(Ljava/lang/String;)Ljava/lang/Integer;"),
        e->GetObjectArrayElement(args,0));
    e->CallVoidMethod(jbs,e->GetMethodID(jcB,"set","(I)V"),
        e->GetIntField(jo,
            e->GetFieldID(jcI,"value","I")));
    jclass jcS = e->FindClass("java/lang/System");
    jclass jcP = e->FindClass("java/io/PrintStream");
    e->CallVoidMethod(e->GetStaticObjectField(jcS,
            e->GetStaticFieldID(jcS,"out","Ljava/io/PrintStream;")),
        e->GetMethodID(jcP,"println","(Ljava/lang/Object;)V"),
        jbs);
}
```

**Figure 2.** Example Java Program Rewritten in Java/C++/JNI

The JNI has its own terminology[6]. An *object reference* has `jobject` type and refers to a Java object (`instanceof Object`). A *class reference* has `jclass` type, is also an object reference, and refers to a Java class object (`instanceof Class`). A *descriptor* is a string value describing the type of a Java field, the signature of a Java method, or the name of a Java class. A *field ID* or *method ID* has either `jfieldID` or `jmethodID` type, respectively, and enables access to Java fields, invoke of Java methods, and construction of Java objects.

# Rationale for Design

Figure 2 shows how overly complex it is to write native C++ using JNI. A great concentration is required to program the native C++ based on what is wanted from a Java perspective. There are too many low level details involving field ID and method ID values, class references, and hard to remember field, method, and class descriptors. It is far better, if to each Java field, method, and constructor, there are available C++ member functions that *completely wrapper*, i.e. *encapsulate* the details.

Skipping to the chase, Figure 3 shows the example of Figure 2 rewritten in a C++ style which encapsulates the JNI calls. This is referred to as *C++ Wrapped JNI* (*C++WJ*).

```
public class Bar {
    public static native void main(String[] a);
    static {System.loadLibraray("foo");}
    };

//C++ from here
#include "Bar.h"  //javah generated
#include "jjava_lang_String.h"
#include "jjava_lang_Integer.h"
#include "jjava_util_BitSet.h"
#include "jjava_lang_System.h"
#include "jjava_io_PrintStream.h"

JNIEXPORT void JNICALL Java_Bar_main(
    JNIEnv* e,jclass this_class,jobjectArray args){
    jjava_lang_StringArray jargs(args);
    jjava_util_BitSet jbs = jjava_util_BitSet::BitSet(e);
    jbs.set(e,jjava_lang_Integer::valueOf(e,
            jargs.GetElement(e,0)).value(e));
    jjava_lang_System::out(e).println(e,jbs);
    }
```

**Figure 3.** Example Java Progam Rewritten in Java/C++WJ

Several C++ classes, *jtypes*, one each for `java.util.BitSet`, `java.lang.Integer`, `java.lang.System`, and `java.io.PrintStream` together offer member functions for C++WJ method invoke, field access, and object construction. Correct JNI function calls with appropriate descriptor parameters encapsulate behind jtype member functions `BitSet`, `set`, `valueOf`, `value`, `out`, and, `println`. An additional jtype for `java.lang.String[]` encapsulates access to JNI function `GetObjectArrayElement` behind member function `GetElement`.

There are three encumbrances compared to the equivalent Java code: 1) wrapped calls require an initial `JNIEnv*` parameter, 2) creating a Java object requires extra syntax involved with a static member instead of a constructor, and 3) Java fields require handling by member calls and not by field access. In other respects the C++WJ code has no more detail and follows along the lines of the corresponding Java code.

A nice benefit of C++WJ is the possibility of compile-time type checking provided the C++ classes have specific enough type information in member signatures. This effectively allows compile time type checking equivalent to the compile time type checking of the Java compiler. This benefit should not be underestimated. Straight JNI calls depend essentially upon run-time type checking which is more developer time consuming and less reassuring toward software quality.

Before moving on to the remainder of the paper, it is pointed out that stretches of JNI code as in Figure 2 are somewhat of a worst case for JNI function calls for one place. Such stretches of code if not larger stretches are often factored with JNI *callbacks* to Java (`Call...Method` function calls) to perform the "heavy lifting". Still, it is not uncommon to have small stretches of JNI function calls in lots of places, so the need for an improved C++ JNI is real. Besides, the need to artificially add relatively small Java methods to classes essentially to circumvent the troubles of JNI usage is not aesthetically appealing. Moreover, a Java class that may make most sense to have such native-callback-to-Java methods may not be amenable to additional methods, for e.g. a 3rd party class.

## Motivating Requirements

C++WJ is motivated by a need to:

- Eliminate the complexity of field ID, method ID, class reference, and descriptor usage.
- Replicate compile-time checking of the Java type system.
- Minimize the performance overhead of repeated JNI use via caching techniques.
- Interoperate with existing "C" level JNI.

The example of Figure 3, if implementable, shows a way to achieve these ends. A C++WJ type is created that contains member functions one for one with the fields, methods, and constructors of the related Java class or interface. The result of calling a C++WJ member function is the appropriate result on the corresponding Java member. A C++WJ class exhibits both the *facade* design pattern[8] against the JNI subsystem as well as the *proxy* design pattern[8] on the related Java class.

C++WJ is an exercise in choosing appropriate C++ constructs for a reasonable C++ style JNI.

## Type Conversion vs. Performance

The contentious part of the technique has to do with balancing type conversion issues with performance. For starters, it should be the case that the three language notions of Java *inheritance*[9]: 1) extending a class, 2) extending one or more interfaces, and 3) implementing one or more interfaces, have appropriate C++ analogues. Then, inherited Java members can likewise become inherited C++ members in the related C++WJ classes. Now, C++ supports only one language notion of inheritance, namely subclassing one of more superclasses, called *multiple inheritance* where more than one superclass is involved. Thus, a Java class or interface reasonably maps or *relates* to a C++ class, and all three Java inheritance forms reasonably relate to C++ multiple inheritance. The usual rules for classes with respect to conversions are satisfied similarly in both Java and C++. That is, subclasses/subinterfaces of a parent class/interface (Java or C++) are allowed in contexts of the superclass/superinterface. Java resolves duplicate *indirect* interfaces of a class (or interface) along different inheritance paths by ignoring the duplicates. C++ public virtual inheritance provides the same ability.

A problem rears up in so far as Java allows casting of superclasses/superinterfaces *down to* subclasses/subinterfaces. An example is the cast required upon getting an element from a `Hashtable` object. The `Hashtable` as a generic container class supports insertion of any type of object and the get operation returns an `Object` object. This returned object must be cast down from `Object` type to some other type in order to use the object in most cases. C++ does not permit casts from superclasses down to subclasses. What C++ does support is `dynamic_cast` of reference to superclass down to reference to subclass provided the complete object's actual type is *polymorphic* and has an unambiguous subclass object. These facts tend to imply the need to use C++ reference types (T&) with some restrictions instead of non-reference types so as to get downcasts in C++WJ. These facts also seem to imply the need for the `dynamic_cast` operator, a relatively expensive operation[11].

C++ reference types (or pointer types which have similar casting properties) are simple enough to use in the arguments to C++WJ member functions, for then the actual objects referred to are outside of the member functions scope and remain after member function return. Reference types are a problem as return type, however, because after member function return, the reference is dangling when the object referred to was created within the member function. Using C++ dynamic memory allocation and requiring programmer memory management is both more complex and too expensive[11] for a task that essentially encapsulates JNI. Requiring a C++ garbage collector with dynamic memory allocation is yet more expensive.

For all of the reasons above, it is decided to avoid C++ reference types and `dynamic_cast` operator and instead to work with non-reference C++ types. The downcast problem is solved by requiring two operations, one a conversion to some prespecified type, followed by a second conversion construction from the prespecified type to the final type. The prespecified type is chosen to be `jobject` type to allow JNI interoperability. As the conversion from prespecified type is available to all C++WJ types, C++WJ effectively supports unchecked Java casts. However, as shown later, C++WJ also supports run-time checked casts. An example of a C++WJ downcast related to the cast of an `Object` to a `Boolean` is shown on the last line in Figure 4.

```
class jjava_lang_Object{//...
    public:operator jobject();
    public:jjava_lang_Object(jobject);
    };
class jjava_lang_Boolean:public virtual jjava_lang_Object{//...
    public:operator jobject();
    public:jjava_lang_Boolean(jobject);
    };

//native implementation... assume jjava_lang_Object jo initialized
    jjava_lang_Boolean jb(static_cast<jobject>(jo));
```

**Figure 4.** C++WJ (Hypothetical) Downcasting

The constructor from and conversion member function to `jobject` type are more than arbitrary. An implementation of a C++WJ type needs access to an object reference to refer to the underlying Java object. Such an object reference as cached via C++WJ type constructor is considered *the* object reference of a C++WJ type object. It does not change value for the lifetime of the C++WJ type object.

The question arises as to what other data members might a C++WJ type class contain and as a result how expensive are C++WJ type constructors. For instance, copy constructor costs are a real issue with parameter passing and return results passed by object values and not object references (or pointers). The expense of C++ copy constructors is related to the expense of base class and member copy constructors and the initialization of implementation specific data. Use of public virtual inheritance can typically add expense to copy constructors for the assignment of hidden implementation data. (Hidden data is required depending upon the implementation and class hierarchy so as to support casting from derived portions of the object to base portions.[11])

By contrast, copy constructor costs can be minimal if extremely lightweight classes are used. As `jobject` type is typically implemented as a C++ pointer, then a class with no superclasses and no virtual member functions and containing only `jobject` type can be no more expensive than a pointer.[10] This is excellent as pointers typically fit in hardware registers. Finally, a single inheritance *hierarchy* of classes containing no virtual member functions and altogether only containing a `jobject`, can be no more expensive than `jobject` type.

Mainly for performance related reasons, C++WJ classes come in two flavors. One flavor, *Jtype* flavor, has public virtual inheritance and potentially non-trivial number of data members and is responsible for full encapsulation of the related Java class or interface. A second flavor, *jtype* flavor, has single inheritance for superclasses, and no more than `jobject` data member and has less than full encapsulation responsibilities. Both flavors have wrapping member functions returning jtype values and taking jtype arguments resulting in potential performance approximating straight JNI calls. Jtypes construct from and convert to jtypes. The jtypes construct from and convert to `jobject` type. The jtypes encapsulate object construction, method invoke, and field access. Via the single inheritance, a jtype makes available direct and indirect superclass method invoke and field access. The jtypes do not make available method invoke and field access inherited from an interface. Instead, a jtype provides a unary postfix operator (`operator++(int)`) to construct the related Jtype, the latter of which has the means to perform all method invoke and field access. The added burden to C++WJ usage is the need to construct Jtype objects out of jtype objects in those places that require interface member visibility. This is an acceptable sacrifice for performance. Finally, both Jtypes and jtypes use JNI name encoding except for a leading "j" or "J" respectively.

On the whole, then, these two flavors of C++WJ classes are reasonable. The new jtypes must now behave with respect to subtyping and downcasts as did the earlier jtypes now considered Jtypes. This must be done without use of multiple inheritance. The solution is for a jtype to use single inheritance to another jtype for the related Java superclass relation, and to use conversion member functions to every direct and *indirect* jtype for the related Java superinterface relation. A jtype related to a Java interface inherits from the jtype related to `Object`. Subtyping conversions are thus in place for jtypes and through a two operations mechanism, downcasts are available for jtypes as in Figure 4.

A subtyping conversion of a jtype to an interface related jtype becomes a second, if slightly less convenient, way to obtain Java interface members. Thus, both `jo++.member()` and `((jInterface)jo).member()` are equivalent for a Java interface `Interface` inherited member `member`. Use of one or the other is required for interface inherited fields of the related Java object or interface inherited methods not declared in the specific type of the related Java object.

There is an alternative design whereby jtypes not only have conversion member functions to all classes related to superinterfaces, but contain the superinterface's members as well. The jtypes still only use C++ single inheritance to classes related to Java superclasses. The jtypes can again have performance similar to `jobject` type. There is also no need for Jtypes as jtypes fully encapsulate a Java class given that member functions for interface members are available directly in a jtype. The problem with this solution is that a change, say addition of members, to a Java interface requires changes to all jtypes whose related class or interface extends directly or indirectly the changed interface. This makes a jtype overly dependent on changes in the Java hierarchy. Member information of super types is better inherited. For this reason, C++WJ types are modestly dependent only upon the related Java type's members and the set of all direct and indirect superclasses and superinterfaces but not upon these Java types' members. This effects a reasonable compromise and matches some of the more important properties of Java binary compatibility.

The C++WJ classes for `Object`, `java.io.DataOutput`, and `java.io.ObjectOutput` as concerns inheritance and special member function declarations are shown in Figure 5.

```
class Object{//...
interface java.io.DataOutput{//...
interface java.io.ObjectOutput extends java.io.DataOutput{//...

//C++ from here
class jjava_lang_Object{//...
    public:operator jobject();
    public:jjava_lang_Object(jobject);
    };
class Jjava_lang_Object{//...
    public:operator jjava_lang_Object();
    public:Jjava_lang_Object(jjava_lang_Object);
    };

class jjava_io_DataOutput:public jjava_lang_Object{//...
    public:jjava_io_DataOutput(jobject);
    public:Jjava_io_DataOutput operator++(int);
    };
class Jjava_io_DataOutput:public virtual Jjava_lang_Object{//...
    public:operator jjava_io_DataOutput()const;
    public:Jjava_io_DataOutput(jjava_io_DataOutput);
    };

class jjava_io_ObjectOutput:public jjava_lang_Object{//...
    public:operator jjava_io_DataOutput()const;
    public:jjava_io_ObjectOutput(jobject);
    public:Jjava_io_ObjectOutput operator++(int);
    };
class Jjava_io_ObjectOutput:public virtual Jjava_io_DataOutput,
    public virtual Jjava_lang_Object{//...
    public:operator jjava_io_ObjectOutput()const;
    public:Jjava_io_ObjectOutput(jjava_io_ObjectOutput);
    };
```

**Figure 5.** C++WJ types for java.io.DataOutput and java.io.ObjectOutput

As Jtypes now do not contain a constructor directly from `jobject` type, but instead from a related jtype, it takes two additional operations to downcast Jtypes over and above the operations to downcast the related jtypes. The native code of Figure 4 is changed to require 4 operations. A more realistic cast, however, is the downcast of jtype to Jtype which takes 3 operations as shown in Figure 6.

```
class jjava_lang_Object{//...
class Jjava_lang_Object{//...
class jjava_lang_Boolean:public jjava_lang_Object{//...
class Jjava_lang_Boolean:public virtual Jjava_lang_Object{//...

//native implementation... assume jjava_lang_Object jo initialized
    Jjava_lang_Boolean jb(jjava_lang_Boolean(static_cast<jobject>(jo)));
```

**Figure 6.** C++WJ (Realistic J/jtype) Downcasting

The conversion of subclass to superclass in C++ is one of the set of standard conversions. Any number of standard conversions can apply before and after a single user defined conversion. This fact allows a Jtype subclass to implicitly convert to a jtype superclass. For example in Figure 7, conversion of JCar to jBarable obtains from standard conversion of JCar to JBarable and then user defined conversion from JBarable to jBarable. Only one dashed line arrow (user defined conversions) is traversed.

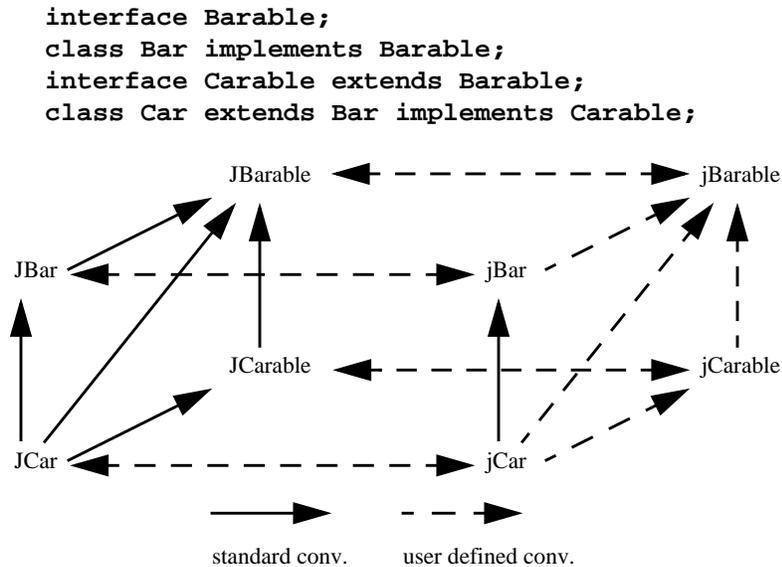

**Figure 7.** Class & Interface Conversions for J/jtypes

Converting from one jtype to a superclass jtype is either a standard conversion for superclasses related to a Java superclass or is a user defined conversion for superclasses related to a Java interface. This implies that interface conversion uses up the one user defined conversion operation of C++ that happens implicitly, but superclass conversion does not. However, a jtype provides conversion member functions to all classes that relate to direct and indirect superinterfaces.Thus, at most one user defined conversion is ever needed for conversions *up the hierarchy*. This is seen in Figure 7, where any lower left class converts to any upper right class crossing at most one user defined conversion (dashed arrows). This is true even though there's a difference between the kind of conversion from jCar to jBar (solid arrow) versus that from jCarable to jBarable (dashed arrow).

The difference between jtype superclass conversion and jtype interface conversion matters for overloaded member functions. Should a function `foo` take each of `jBarable`, `jCarable`, and `jBar` of Figure 7 in three different overloaded functions, then a use of `foo` on a `jCar` object results in C++ compiler choice of `foo(jBar)`. This is due to the C++ best matching function algorithm having preference for standard conversions over user defined conversions. At the same time, take away `foo(jBar)`, and the C++ compile fails for an ambiguity between `foo(jBarable)` and `foo(jCarable)`. This is true even though `jCarable` type is closer (direct Java interface) to `jCar` type than `jBarable` is (indirect Java interface). The design of jtypes thus sacrifice a knowledge of proximity of interfaces in order to minimize copy costs. This is not considered much of a drawback, however, as the issue manifests only when Java interface members are overridden. And from a Java standpoint, overridden interface members are simply a redundancy for method members and somewhat of a poor practice for data members.

## Wrapped Calls

C++WJ types have the responsibility for providing JNI method invoke, field access, and object construction. C++WJ type member functions corresponding to operations on each Java member are therefore provided. For example, setting a Java object field is accomplished with a call to a C++WJ type member function taking `JNIEnv*` and a jtype value. Getting a Java object field is accomplished with a call to a C++WJ type member function taking `JNIEnv*` and returning a jtype value. So a Java field corresponds to two C++WJ type member functions of different signatures, one to set the value and one to get the value. A Java `final` field, however, corresponds to only the get member function as a `final` field value can not be changed.

Each Java method and constructor corresponds one to one to a C++WJ type member function. Java class (i.e. `static`) fields and class methods correspond to C++WJ type `static` member functions. Overloaded Java methods, as one could expect, correspond to overloaded C++WJ type member functions. A Java method and field with the same name corresponds to overloaded C++WJ member functions taking an extra initial parameter of differing types.

Each Java type requires a distinguishing C++WJ type to allow C++WJ type member function overloading. A JNI primitive type such as `jint` suffices and implements properly for its related Java type. Java classes and interfaces are handled by their related jtypes. Finally, by extension, C++WJ provides jtypes for arrays, as for example `jjava_lang_IntegerArray` for `Integer[]`. The implementation of jtypes for Java arrays is described in the next section.

The term *jtype*, henceforth, applies to all these C++WJ types and the term *class jtype* refers to jtypes related to Java classes and interfaces only. Table I exemplifies the different kinds of Java types and related C++WJ jtypes and implementation.

| Java Type | C++WJ jtype | C++WJ jtype implementation |
|---|---|---|
| void | void | C++ primitive type |
| int | jint | JNI primitive type |
| int[] | jintArray1 ("1" needed) | C++ class |
| int[][] | jintArrayArray | C++ class |

**Table I.** Prototypical Native Types

| Java Type | C++WJ jtype | C++WJ jtype implementation |
|---|---|---|
| java.lang.Integer | jjava_lang_Integer | C++ class |
| java.lang.Integer[] | jjava_lang_IntegerArray | C++ class |

**Table I.** Prototypical Native Types

With a complete set of jtypes, C++WJ type wrapper member functions can correspond to all cases of Java field types, method parameter and return types, and constructor return (class) types. Example C++WJ jtype wrapper member functions for all corresponding field, method, and constructor members of `Boolean` are shown in Figure 9.

```
class jjava_lang_Boolean:public jjava_lang_Object{//...
    public:inline static jjava_lang_Boolean TRUE(JNIEnv*);
    public:inline static jjava_lang_Boolean FALSE(JNIEnv*);
    public:inline static jjava_lang_Class TYPE(JNIEnv*);
    public:inline jboolean value(JNIEnv*);
    public:inline void value(JNIEnv*,jboolean);
    public:inline static jlong serialVersionUID(JNIEnv*);
    public:inline static jjava_lang_Boolean Boolean(JNIEnv*,
        jboolean);
    public:inline static jjava_lang_Boolean Boolean(JNIEnv,
        jjava_lang_String);
    public:inline jboolean booleanValue(JNIEnv*);
    public:inline static jjava_lang_Boolean valueOf(JNIEnv*,
        jjava_lang_String);
    public:inline jjava_lang_String toString(JNIEnv*);
    public:inline jint hashCode(JNIEnv*);
    public:inline jboolean equals(JNIEnv*,jjava_lang_Object);
    public:inline static jboolean getBoolean(JNIEnv*,
        jjava_lang_String);
    public:inline static jboolean toBoolean(JNIEnv*,
        jjava_lang_String);
};
```

**Figure 8.** jtype Field, Method, and Constructor Declarations for Boolean

A reasonable question to ask is "Why doesn't a Java constructor correspond to a C++WJ type constructor?". There are a few reasons. The best reason has to do with interoperability with JNI "C", in that C++WJ types must already construct from either class jtypes for Jtypes or from `jobject` type for jtypes. These constructors cannot create a new Java object (nor even an additional object reference) but must simply encapsulate the existing object reference. Thus, there is already very general C++WJ type constructors in place. Allowing for additional C++WJ type constructors which both create and encapsulate a Java object is possible but not so desirable. There is a potential for confusion in the case of a Java constructor taking a same typed parameter. A C++WJ type constructor for this Java constructor would differ from the C++WJ type default copy constructor by only the JNIEnv* initial parameter. A second problem is that a constructor for a Java class with superclasses leads to a C++WJ type constructor required to call some C++WJ type superclass constructor essentially to do nothing. A do-nothing constructor for C++WJ types would be then be needed as the C++WJ default constructor does not work given that it initializes the object reference to 0. Overall, it is desirable not to combine Java object construction with C++WJ type object construction.

The C++WJ type member function corresponding to a Java constructor is necessarily a `static` member function. Because of the leading extra "j" or "J" in a C++WJ type name, a C++WJ type member function corresponding to Java constructor does not clash with C++WJ type constructors. A package name also adds to the distinction as it is part of the C++WJ type name but not part of a C++WJ type member function name (unless it was the case in the Java class).

## Conversions Revisited

C++WJ type wrapped calls present Java field access, method invoke, and object construction through C++ member functions. The C++ member function parameter and return assignments should reasonably behave so as to mimic Java implicit conversions at method parameter pass, method return assignment, and field assignment. In Java and with one exception for assignment, the conversions involved are the *widening* conversions, both *widening primitive conversions* and *widening reference conversions*. The Java conversion of `byte` to `short` is an example of a widening primitive conversion. The Java subclass-for-a-superclass conversion is an example of a widening reference conversion. The one exception for Java assignment is the conversions of `int` to `byte`, `short`, or `char` which are *narrowing* conversions. C++WJ needs to handle corresponding widening and narrowing conversions for jtypes.

The jtype conversions corresponding to Java widening primitive conversions are handled by a combination of the JNI primitive type implementation and C++ primitive conversions. For example, `jbyte` converts to `jshort`, as both are JNI implemented as appropriately sized C++ integral types and the conversion is a C++ *integral promotion*. The jtype conversions corresponding to Java widening reference conversions are handled by the design of C++WJ jtypes. Class jtypes are taken care of by subclass-for-superclass conversions already considered. Java array (to any dimension) of class related jtypes are handled by requiring such jtypes to have conversions matching that of the base types involved. This supports Java array-of-subclass-for-array-of-superclass conversion. The array jtypes also have conversion to `jjava_lang_Object` and `jjava_lang_Cloneable` to support corresponding Java reference conversions.

As there are an infinite number of array dimensions, array to any dimension of either primitive jtypes or class jtypes reasonably implement with C++ templates. As the need for a particular array jtype of a particular dimension is needed, a C++WJ template taking the `unsigned int` dimension instantiates. The first two dimensions are afforded simpler usage with `typedefs` to *jtype*Array or *jtype*ArrayArray identifiers.

The jtype declarations for `Boolean[]`... are shown in Figure 9.

```
template<unsigned int n>
class jjava_lang_BooleanARRAYD:public jjava_lang_Object{//...
    public:inline operator jjava_lang_ObjectARRAYD< n >()const;
    public:inline operator jjava_io_SerializableARRAYD< n >()const;
    public:inline operator jjava_lang_Cloneable()const;
    public:inline jjava_lang_BooleanARRAYD();
    public:inline jjava_lang_BooleanARRAYD(jobject);
    };

typedef jjava_lang_BooleanARRAYD< 1 > jjava_lang_BooleanArray;
typedef jjava_lang_BooleanARRAYD< 2 > jjava_lang_BooleanArrayArray;
```

**Figure 9.** C++WJ types for Boolean[]...

There is an alternative array jtype implementation using element jtype as a template parameter. One advantage is that a single ARRAY template can handle all jtypes at least up to a fixed number of member conversions to other array jtypes. The problem with this approach is that each member conversion to array jtype requires its own element type and one that is different than the template element type. Hence, there is a need for additional template arguments for these other element types, if not for the array jtypes themselves. The higher dimensional usage also requires nested template instantiations that become overly complicated on the additional arguments. The simpler approach using templates taking unsigned int is preferred even if one template per JNI primitive and class jtype is needed.

With the array conversions in place, only the *narrowing* conversions for Java assignments are left. The JNI implementation of jint, jbyte, jshort, and jchar as integral types and C++ support of conversion among integral types covers this case. What happens in C++ in case an input integral value cannot be represented in the target integral type is left implementation defined. Typically, the value is truncated in some fashion.

## Wrapper Implementation

The implementation of a C++WJ type member functions is straight forward using JNI functions except as regards caching (performance) issues. Table II summarizes the cases for jtypes. The implementation of a Jtype member simply calls the corresponding jtype member as the default. In general, for each jtype, the C++WJ implementation follows known technique to cache a class reference using JNI functions FindClass followed by NewGlobalRef (or NewWeakGlobalRef in JNI 1.2)[6]. (A *global* or *weak global* object reference is required for use across native invocations.) The result is cached in a jtype private static data member. The FindClass is asterisked as calling it may not be required as discussed in the next section. This cached class reference is considered *the* (current) class reference of a C++WJ type (jtype as well as Jtype). The C++WJ implementation follows a second known technique to statically cache a field ID or method ID value based on the cached class reference.[6]

| Java member | JNI function to use | JNI functions to use for cache |
|---|---|---|
| Constructor | NewObject | FindClass*,GetStaticMethodID |
| Static field | Get/SetStatic...Field | FindClass*,GetStaticFieldID |

**Table II.** C++WJ jtype Implementation Cases

| Java member | JNI function to use | JNI functions to use for cache |
|---|---|---|
| Non-static field | Get/Set...Field | FindClass*,GetFieldID |
| Static method | CallStatic...Method | FindClass*,GetStaticMethodID |
| Non-polymorphic method | CallNonvirtual...Method | FindClass*,GetMethodID |
| Polymorphic method | Call...Method | FindClass*,GetMethodID |

**Table II.** C++WJ jtype Implementation Cases

An overridden non-static non-final non-private method or a method from an abstract class or interface, i.e. a *polymorphic method*, is implemented with a superclass or interface class reference computed method ID value and does not require a method ID value computed from a class reference of an invoking object's most specific type (JNI function GetObjectClass on an object reference).[7] It is enough to use the method ID value computed from the cached class reference of a C++WJ type. The JNI subsystem, itself, manages to invoke the correct polymorphic Java method. Furthermore, the C++WJ implementation use of a field ID value also computed from a class reference is essential in the case of overloaded fields. Use of a field ID value computed from run-time class reference would lead to an inappropriate field access.[6]

A C++WJ type does not attempt, in general, to cache the value of Java fields so as not to invite stale cache problems between a C++WJ type object and its related Java object for even the briefest of running time. Java final fields, however, are amenable to caching given that a final field value cannot be changed after initialization. A final class (i.e. static) field is cached as a C++ static data member. A final static object field is cached as a global reference (weak global reference in JDK 1.2).

A final instance (i.e. non-static) field, if cached, would have to be cached as a C++ non-static data member. A Jtype at greater expense to its copy constructor cost therefore caches final field values, given that jtypes cannot afford additional instance fields. A final instance object field is cached only if the object reference is not a global reference and not a weak global reference. Otherwise, it would be necessary to cache a final instance object field as a global or weak global reference. But then, the Jtype destructor has to call JNI DeleteGlobalRef or DeleteWeakGlobalRef on these cached global or weak global references. This requires a *valid* JNIEnv*. But a Jtype does not cache a JNIEnv* as these are thread specific. As destructors do not take arguments, there is no way to pass in a valid JNIEnv*. A Jtype would need access to the JNI JavaVM* to realistically obtain the JNIEnv* via JNI function AttachCurrentThread. All of this ignores the further problem of which Delete...Ref to call. Thus, it is not feasible to cache final object fields for a global or weak global object reference. Moreover, if it is not possible to determine if the object reference is not a global or weak global reference (JNI implementation specific), then it is not possible to cache *any* final instance fields. (Some JNI implementations distinguish global and weak global references from local references by negative as opposed to positive int cast from jobject values.)

Assuming it possible, the C++WJ implementation set up for caching of final instance fields for a *local* object reference does not adversely affect performance of a Jtype destructor or constructor or else is not undertaken. For example, a reasonable C++WJ implementation may decide that if the number of final instance fields in a Java class is greater than the number of bits in a C++ int, then no final instance field caching is undertaken. Otherwise, one C++ int data member initialized to 0 in the Jtype constructor taking jtype and treated as a bitset relatively quickly indicates all final instance field caches are empty. A more robust implementation uses as many C++ ints as needed and has no limit on the number of final instance fields that can be cached.

The implementation of C++WJ wrappered calls implements in an `inline` fashion for the best possible performance. Static values caching, however, is not expected to occur more often than a related Java class is loaded, so can implement in a non-`inline` fashion. The C++WJ implementation performs conversions between jtypes and `jobject` and `jclass` types everywhere as needed.

## Caching the Class Reference

As mentioned in the previous section, C++WJ type member function implementations use `FindClass` assigned class reference. The JNI function `FindClass` in JDK 1.2 JNI uses the class loader that defined the class whose native method triggered a native invocation. As a given Java class used in a `native` method implementation may be loaded by some other class loader, then it is required that C++WJ allow for explicit caching of the class reference. An alternative requiring C++WJ to instead lookup and assign a class reference from methods on a `ClassLoader` object adds assurance but sacrifices considerable performance. Moreover, `static native` implementations via JNI passed `jclass` parameter can take advantage of explicit class reference caching. This saves the C++WJ implementation a call of `FindClass`. Finally, explicit caching of a class reference to 0 has the added benefit of releasing cached global references (weak global references in JDK 1.2). This can be important to an application.

Explicit caching of a class reference is provided by a jtype `static` member function taking a `jjava_lang_Class` parameter. The name of this `static` member function has to be distinct from all C++WJ type member functions corresponding to Java members. A decent, if not meaningful, choice is the Java keyword `native`. This choice is guaranteed both not to conflict with member names corresponding to Java members and also to compile as a C++ member name.

Figure 10 shows a `static native` method `nativeInit` calling C++WJ `native(jjava_lang_Class)` member function. The `nativeInit` method is invoked in a `static` block to perform caching at class load (and reload) time.[6]

```
public class Bar{//...
    private static native void nativeInit();
    static{System.loadLibrary("foo"); nativeInit();}
    };

//C++ from here
JNIEXPORT void JNICALL Java_Bar_nativeInit(JNIEnv *e,
    jclass this_class){
    jBar::native(e,jjava_lang_Class(this_class));
    }
```

**Figure 10.**  First Static Block Invoking Native Initialization

The jtype member function `native(jjava_lang_Class)` follows through and performs all field ID and method ID value caching and all `final static` field caching besides caching the class reference. However, even with the availability of the member function `native(jjava_lang_Class)`, the C++WJ implementation may find it has not been called and be forced to implicitly call `FindClass`. As this case often happens post class load time when performance costs are typically more important, the C++WJ implementation does not cache beyond the class reference when `FindClass` is used. Instead, these further values are cached as needed at each and every first use. C++WJ type wrapped calls must therefore expend a minimal check to see if field ID or method ID value caching or `final static` field caching has happened or not and if not whether `jclass` caching has happened or not. The extra cost of these checks is minimal against the at least one JNI function call that occurs.

A class jtype member function `native(jjava_lang_Class)` calls its superclass' member function `native(jjava_lang_Class)` to perform caching of values of the related Java superclass. The class reference of the superclass is obtained from JNI function `GetSuperClass`.

**Type Checking**

A reasonable C++WJ type implementation as outlined uses jtypes in wrapped call arguments and return values. It has been shown that jtypes related to Java class or array of class satisfy subclass-for-a-superclass conversion through a combination of single inheritance and conversion member functions. It also has been shown that downcasting works for a jtype related to Java class or array of class through a couple of conversions. The need to provide both downcasting and interoperability to `jobject` type leads, however, to the fact that any non-primitive jtype converts to *any* other non-primitive jtype via two user defined conversions.

The fact that C++ allows only *one* user defined conversion operation implicitly together with a C++ compiler's type checking provides C++WJ usage a solid measure of type checking. This type checking ability ought not be down-played as it provides assurance before run-time that the native code usage of the underlying JNI is sensible. Un-wrapped JNI mainly manipulates `jobject` type and so development with straight JNI functions cannot rely on any serious compile time type checking. By contrast, a compiler caught type error in C++WJ using `jjava_lang_Boolean` of Figure 9 is shown in Figure 11.

```
//native implementation...
//assume jjava_lang_Boolean jb and JNIEnv* e initialized
    jb.valueOf(e,jb);//err: no jjava_lang_Boolean to jjava_lang_String
```

**Figure 11.** Compile Time Type Checking Error on Bad Argument Type

C++ implicit conversions apply at parameter passing and returning a result expression, but not at the invoking object expression. This fact along with the user defined conversion of `jobject` type to any class jtype implies that `jobject` can be passed as an object argument to any wrapper member function without compile error. However, an object of `jobject` type cannot be used as invoking object to a wrapper member function without error. The conversion to a C++WJ type must be explicit at the invoking object.

One area not receiving complete type checking support is primitive implicit conversions. C++ allows narrowing conversions beyond those of similar Java types, for example C++ `float` to `byte`. C++WJ is not capable of type checking these conversions as performance dictates `jfloat` and `jbyte` as implemented in JNI typically as typedefs of C++ primitive types.

Finally, to discourage direct use of C++WJ conversion to `jobject` type, a separate C++WJ macro `JNICAST(/*JNIEnv**/e,/*jjava_lang_Object*/o,/*jtype*/t)` is provided. The macro body of `JNICAST` in addition to providing appropriate C++WJ conversion, also provides Java run-time conversion checking by invoking JNI function `IsInstanceOf` in such a way as to return a null object if the object argument is not convertible to the jtype. A C++WJ user application can further wrapper this macro call on null return to throw a C++ exception or call error handling as desired.

## Placeholder Interfaces

Interfaces such as `Cloneable`, `java.io.Serializable`, and `java.rmi.Remote` which have no fields or methods typically serve a general semantic (even language level) purpose. As these interfaces are pervasive and do not require C++WJ wrapped access, then C++WJ does not have Jtypes for them. At the same time, these interfaces can be arguments, return values, and field types in Java. Thus, C++WJ has class jtypes for them. Jtypes and class jtypes related to Java classes (or interfaces) that implement (or extend) such placeholder interfaces still require user defined conversions to the class jtypes related to these interfaces.

## Complete Wrappering of JNI

Besides method invoke, field access, and object construction, JNI provides additional functions related to JNI functions involving `jstring`, `jthrowable`, `jclass`, `jobject`, `jobjectArray`, and primitive array types. C++WJ has to convert to/from related jtypes such as `jjava_lang_String` for `jstring` and/or has to wrapper these JNI functions. It is more consistent to wrapper the JNI functions with extra jtype member functions. This way the C++WJ application programmer does not need both kinds of types and can work consistently with C++WJ jtypes.

An example `jjava_lang_Object` wrapper member function for JNI function `IsSameObject` is seen in Figure 12. The implementation is straight forward.

```
class jjava_lang_Object{//...
    public:inline jboolean IsSame(JNIEnv*,jjava_lang_Object);
    };
```

**Figure 12.** Convenience Member IsSame for jjava_lang_Object

JNI functions involving primitive array types or `jobjectArray` type are wrapped by member functions within array jtypes. For example, `jjava_lang_ObjectArray::GetLength` returns the length of an `Object` array. The C++WJ member function `New` wrappering JNI function `newObjectArray` for a particular array jtype uses knowledge of the element type (one fewer dimension) and so does not need a wrapped class reference parameter. The one dimension array jtype case uses the class jtype's cached class reference. As an example, the convenience member functions for `jjava_lang_ObjectARRAY< n >` are shown in Figure 13.

```
template<unsigned int n>
class jjava_lang_ObjectARRAYD:public jjava_lang_Object{//...
    public:inline jsize GetLength(JNIEnv* e)const;
    public:inline jjava_lang_ObjectARRAYD< n-1 > GetElement
        (JNIEnv* e,jsize index)const;
    public:inline SetElement(JNIEnv* e,jsize index,
        jjava_lang_ObjectARRAYD< n-1 > value);
    public:inline static jjava_lang_ObjectARRAYD< n >
        New(JNIEnv* e,jsize length,
        jjava_lang_ObjectARRAYD< n-1 > initialElement);
    };
```

**Figure 13.** Convenience Member Functions for jjava_lang_ObjectArray[]...[] (n>1)

The use of the element type (template instantiation at `n-1`) in the array jtype template convenience member functions forces a need for template specialization at dimension `n=0` so as not to template instantiate at `0-1`. There are specializations also at `n=1` to reference the appropriate C++WJ element type. The `n=1` case for primitive array jtypes is fully implemented to wrapper all JNI primitive array functions.

The class reference for an array jtype is cached in a `static` data member. The `New` member function at next higher dimension requires access to the cached class reference. An array jtype's class reference becomes invalid should the underlying class jtype's class reference change. (Say, the underlying class jtype's class reference is reset with `native` member function to reference a same named class from a different class loader.) To handle this, the `native` member function of a class jtype locates and resets to 0 the cached class references of all array jtypes with it as base type. One implementation for this uses a linked list of address of class references with list head information stored with the class jtype. After an array jtype class reference is reset to 0, the next use of the class reference causes a recomputation. The recomputed class reference is obtained in general from combined calls of `GetObjectClass` of `NewObjectArray` using already computed one lower dimension class reference. An alternative implementation uses C++WJ wrapped invoke of `findClass` from a cached `jjava_lang_ClassLoader` object that is the class loader of the underlying class jtype. For JDK 1.2, if a `jjava_lang_ClassLoader` object is not obtainable due to `SecurityException`, then the first implementation is applied. Ordinarily, the first implementation is preferable given that it requires creation of a Java array object rather than a `String` object and array object creation is a considerably faster JVM operation. (An enhanced JNI `FindClass` function taking an object reference to a class loader object along with C/C++ string would make the second implementation preferable by eliminating the `String` object creation.) Either way ensures that caching an array class reference has involved the same correct class loader as that involved with caching the underlying class jtype's class reference.

## Reflection Support

At times, perhaps for debugging or perhaps to make efficient use of JDK 1.2 JNI reflection functions[16], C++WJ needs to provide class references, and field ID and method ID values. C++WJ types have a `static` member function per corresponding Java field, method, or constructor. The member functions take one real `JNIEnv*` pointer and one dummy `JNIEnv*` pointer and return either a field ID or a method ID value, respectively. The dummy `JNIEnv*` parameter declaration is needed as the return type is not

C. Bordelon - 19sufficient to distinguish against a C++WJ wrapper member function of the same arguments. The value returned is the cached field ID or method ID value or a newly computed and cached value as necessary.

The JNI reflection functions `ToReflectedField` and `ToReflectedMethod` require a class reference, so C++WJ reflection support has to provide it. A C++WJ jtype `static` member function `native` taking just `JNIEnv*` parameter returns the current class reference wrapped as a `jjava_lang_Class` object.

Reflection support member function declarations for `jjava_lang_Boolean` are shown in Figure 14. A numerically increasing suffix is added to reflection support members corresponding to overloaded Java members. Which overloaded member gets which suffix follows the order of `Method` elements from an invoke of `getDeclaredMethods` on appropriate `Class` object.

```
class jjava_lang_Boolean:public jjava_lang_Object{//...
    public:inline static jjava_lang_Class native(JNIEnv*);
    public:inline static jfieldID TRUE(JNIEnv*,JNIEnv*);
    public:inline static jfieldID FALSE(JNIEnv*,JNIEnv*);
    public:inline static jfieldID value(JNIEnv*,JNIEnv*);
    public:inline static jfieldID serialVersionUID(JNIEnv*,
        JNIEnv*);
    public:inline static jfieldID TYPE(JNIEnv*,JNIEnv*);
    public:inline static jmethodID Boolean(JNIEnv*,JNIEnv*);
    public:inline static jmethodID Boolean_2(JNIEnv*,JNIEnv*);
    public:inline static jmethodID booleanValue(JNIEnv*,JNIEnv*);
    public:inline static jmethodID valueOf(JNIEnv*,JNIEnv*);
    public:inline static jmethodID toString(JNIEnv*,JNIEnv*);
    public:inline static jmethodID hashCode(JNIEnv*,JNIEnv*);
    public:inline static jmethodID equals(JNIEnv*,JNIEnv*);
    public:inline static jmethodID getBoolean(JNIEnv*,JNIEnv*);
    public:inline static jmethodID toBoolean(JNIEnv*,JNIEnv*);
    };
```

**Figure 14.** C++WJ jtype Reflection Support Declarations

Another use for C++WJ reflection support member functions is to call instance methods of a superclass. There is no direct support for Java `super.m()` type calls in C++WJ, though JNI supports this with JNI `CallNonvirtual...Method` functions. However, Figure 15 shows a native implementation invoke of the `Object::equals` method using C++WJ reflection member functions. The `native(e)` value is not explicitly converted to a `jclass` value as `jjava_lang_Class` type provides this conversion. The last parameter to `CallNonVirtualBooleanMethod` is explicitly converted to `jobject` type as a conversion is not applied to a variable argument list parameter.

```
public class Bar {
    public native boolean equals(Object b);
    static {System.loadLibraray("foo");}
    };

//C++ from here
#include "jBar.h" //includes javah generated Bar.h
#include "jjava_lang_Object.h"

JNIEXPORT jboolean JNICALL Java_Bar_equals(
    JNIEnv* e,jobject this_obj,jobject obj){
    jBar jthis(this_obj); jjava_lang_Object jo(obj);
    return e->CallNonvirtualBooleanMethod(jthis,
        jjava_lang_Object::native(e),
        jjava_lang_Object::equals(e,e),static_cast<jobject>(jo));
    }
```

**Figure 15.** Direct Invoke of Object::equals from Bar Object

### Native Considerations

Calling a `native` method from native code is done in one of two ways: 1) use JNI (or C++WJ) to invoke the method as for any Java method, or 2) directly call the `Java_` function that is the `native` method implementation provided it is accessible from the calling code. (A use of the **javah** tool provides `Java_` function declarations for all native methods.) The second case above is also wrappered in C++WJ. As an option C++WJ provides for each Java `native` method to have a convenience C++WJ type member function taking an initial `JNIEnv&` followed by the same jtype arguments as the associated wrapper member function. The implementation encapsulates a call to the `Java_` function. Figure 16 shows an example using C++WJ to directly call a native method implementation which in this case leads to a *very* long running program.

```
public class Bar {
    public static native void main(String[] a);
    static {System.loadLibraray("foo");}
    };

//C++ from here
#include "jBar.h" //includes javah generated Bar.h
#include "jjava_lang_String.h"

JNIEXPORT void JNICALL Java_Bar_main(
    JNIEnv* e,jclass this_class,jobjectArray args){
    jBar::main(*e,jjava_lang_StringArray(args));
    }
```

**Figure 16.** C++WJ Directly Called Native Method

Wrappering the `Java_` function itself raises the issue of whether the `Java_` function could instead implement as a jtype member function. This would require change to the JVM essentially to check either for `Java_` function or for jtype member function. Also, for jtype member functions, the JVM would have to construct jtype arguments and handle jtype return values and not provide for `jstring, jthrowable, jobject,` or `jobjectArray` arguments and return values. The benefit would be that the native implementation would not have to convert the native function object arguments to start using C++WJ.

## Errors

JNI functions can return error values. There is a corresponding Java exception which through JNI `ExceptionOccurred` can be obtained in native code. The C++WJ implementation checks for error on return from all JNI functions used. If an error has occurred, then a C++ exception object of type `JNIFailure` is thrown. This seems to be a reasonably safe choice. The example in Figure 10 rewritten to properly catch C++WJ thrown JNI errors is shown in Figure 17.

```
public class Bar{//...
    private static native void nativeInit();
    static {System.loadLibrary("foo"); nativeInit();}
    };

//C++ from here
JNIEXPORT void JNICALL Java_Bar_nativeInit(JNIEnv *e,
    jclass this_class){
    try { jBar::native(e,jjava_lang_Class(this_class));
        }
    catch(JNIFailure) {
        //error handling or return to allow handling by caller
        }
    }
```

**Figure 17.** C++WJ Exception Checking JNI Errors

`JNIFailure` error objects carry no information, but also do not require any, as JNI assures that an `ExceptionOccurred` object is available (and `ExceptionCheck` in JDK 1.2 returns true) after all detectable JNI function errors. This includes JNI functions returning error codes and those unable to.[6]

The decision to throw a C++ exception on JNI errors means that failing to `try`/`catch` C++WJ/JNI errors terminates the application via termination handler. On the plus side, C++WJ eliminates the majority of errors by the compile time checking. What's left tends to be explicit casting errors and issues involving classes that are not loadable. An earlier C++WJ design left out JNI error checking altogether, figuring that JNI errors are unlikely, especially in compile environments that regenerate C++WJ output as Java classes change. This is now changed so that applications can be comfortable that errors are not disregarded. C++WJ wrapper member functions are not implemented with C++ exception specification (`throw` clause). C++ `throw` clauses are unchecked at compile time unlike Java `throws` clauses. As the C++WJ implementation has no specific exceptions to indicate, there is relatively little to gain against a possible run-time performance cost to using the C++ exception specification for C++WJ member functions. Finally, the decision to check for and throw C++ exceptions within the C++WJ implementation add somewhat to execution costs. The quality of a given C++ compiler's optimized handling of C++ exceptions in the no-exception-thrown execution case mostly determines the cost overhead.

# jH and C++ Compile

The scope of the C++WJ implementation forces a need to make it as convenient as possible to use. For this reason, a tool, here named *jH*, generates C++WJ class code into C++ header (`#include`-able) output files.

## C++ Source Outputs

**jH** output separates declaration from implementation as much as possible. **jH** outputs C++ `static` data and non-`inline` member function definitions within C++ templates. This allows **jH** not to have to generate source as opposed to header output files and instead relies upon the compiler to *instantiate* C++WJ external definitions. The jtype template output for Java array of classes/interfaces is included with the jtype output for the underlying class. The jtype templates related to Java *primitive* arrays are output with the jtype output for `java.lang.Object`. **jH** has a JNI vendor specific (non-portable) option to indicate that Java `final` instance fields be cached. **jH** has an option to output the direct wrapper member function for native methods. Finally, much of the **jH** generated implementation is the same from class to class and is factored out into C++WJ macros, `inlines`, `templates` and such, so as to simplify the **jH** output.

**jH** is implemented in Java using reflection as one possible implementation. The JDK 1.1 implementation unlike the 1.2 implementation has one caveat as a result of JDK 1.1 `Class.forName` not having an option to suppress class initialization. JDK 1.1 **jH** run for a class executes `static` initializers which in turn attempts to execute native code which is not yet compiled waiting on the output of **jH**. Thus, **jH** generation of C++WJ as in the example in Figure 10 is worked around by either supplying a dummy native library not needing **jH** outputs or temporarily commenting out `native` method invocations from the `static` initializer until **jH** outputs are created.

The **jH** output `#include`'s **jH** output of other Java classes as necessary. The `#include`'d outputs contain the C++WJ types related either to Java superclasses or superinterfaces or to Java parameter, return, or field types. Thus, one invocation of **jH** potentially requires invocations of **jH** for a plethora of other Java classes, or otherwise the result cannot compile. The advice is either to run **jH** for all output files as the need arises or to use the **jH** *-r* (recursive) option to have **jH** automatically perform this. The former is preferred in compilation environments with automated build dependency utilities such as *nmake*[17] as minimal regeneration and subsequent rebuild of **jH** output results. In addition, **jH** has a *-thin* option to not generate wrapper member functions nor Jtypes. This option results in less need of other **jH** output and so is used for all *incidentally* encountered Java classes (classes encountered which are not a superclass or superinterface, but only member involved). **jH** used without *-thin* handles in a *-thin* manner all incidentally encountered Java classes.

The **jH** output is compiled and the result becomes part of a native library. If the same compiled result of **jH** output becomes part of multiple native libraries all loaded into a Java application, then no harm results. The instantiated external definitions are either text or `static` data. The changeable `static` data are cached values and more than one cache is not a problem.

**jH** retains the options *-private*, *-protected*, and *-public* that **javah** has. This makes it possible to restrict or enlarge the output of **jH** and offer some degree of access control. The defaults are the same as **javah**. It is necessary to use *-private* to get `private` member output, for example. In general, **jH** is not gratuitously incompatible with **javah**. This makes it conceivable that **jH** functionality can be combined with **javah** functionality in a single tool.

## Filtering Out Compile Conflicts

**jH** can generate C++ names that may not compile in a C++ environment. A good example is the `jjava_lang_Boolean` member functions `TRUE` and `FALSE` corresponding to fields of `Boolean`. In some compile environments these names clash with C/C++ preprocessor macro names. Even a source file with a single `#include` of a **jH** header output file sometimes does not compile because **jH** output `#include`'s JDK jni.h which in turn can `#include` compile environment header files. Also, **jH** generated names beginning with "J" or "j" can clash either with other such names (class name vs. member name) or even with JNI type names (for example, `JNIEnv`, `jmethodID`, ...). **jH** generated output can even clash with C++ keywords, for example `unsigned`. For all these cases, a reasonable approach is to require separate filtering of **jH** output to replace clashing names with non-clashing names. **jH** output carefully puts names that must not be renamed inside double quote (") marks everywhere.

Other than the above possible clashes, C++WJ has made reasonably sure that compile issues do not come up. As mentioned above, the name `native` is chosen for a certain C++WJ type member name as it can not clash in C++WJ output.

# Suggestive Performance Results

A simple performance comparison of the cpu execution times of the programs in Figure 2 and Figure 3 was done using JDK 1.1 JNI on stock RISC server hardware and a prototype **jH**. The programs were compiled with the vendor C++ compiler using the compiler optimization setting. The programs were slightly modified to invoke the original `main` method within a `for` loop from a new non-`native` `main`. Outputs were sent to a null device. Program executions of 100,001 iterations through the loop and one iteration through the loop were performed. The program cpu execution time difference was used as a preliminary result. The subtraction of the one iteration program execution cpu time eliminates JVM start up cost, initial class loading cost, and one time out-of-cache computation costs. This amount was approximately 1% of the best times. The Java loop and entry/return from native code was included in the costs and was approximately .5% of the best times. The execution was performed 12 times with removal of the single best and single worst runs. The average of the remaining 10 executions was used as a result.

The results showed that the program in Figure 3 surrounded by `try/catch` on error executed in approximately 20% less cpu execution time than the program in Figure 2. This is largely due to the 12 (out of a total of 18) fewer JNI function calls each iteration after the initial iteration. Hand optimization using caching techniques of the code in Figure 2 along with `try/catch` on error resulted in a program with cpu execution time approximately .5% better than that of the cpu execution time of the program in Figure 3 with `try/catch`. On each iteration after initial iteration, the hand-optimized `try/catch` Figure 2 version executed the same 6 JNI calls as did that of the `try/catch` Figure 3 version. On each iteration after initial iteration and not counting JNI function execution, the hand optimized `try/catch` Figure 2 version executed 92 RISC instructions versus 109 for that of the `try/catch` Figure 3 version. Commenting out the `try/catch` and all JNI function error checking of the hand optimized Figure 2 version brought the number of RISC instructions executed down to 79 (not counting JNI function execution). Finally, moving the caching code to a separate `static` block invoked native implementation and removing all checks as to whether data is in cache as shown in Figure 18 brought the number of RISC instructions executed down to 57 (not counting JNI function execution). This best hand-optimized program execution time was approximately 1% better than that of the cpu execution time of the program in Figure 3 with `try/catch`.

```
public class Bar{
    public static native void main(String[] a);
    private static native void otherInit();
    static {System.loadLibrary("foo"); otherInit();}
    };

//C++ from here
#include "Bar.h" //javah generated
static jclass jcB = 0, jcI = 0, jcS = 0, jcP = 0;
static jfieldID jfV = 0, jfO = 0;
static jmethodID jmB = 0, jmV = 0, jmS = 0, jmP = 0;
static jobject jout = 0;

JNIEXPORT void JNICALL Java_Bar_otherInit
    (JNIEnv *e, jclass this_class){
    jcB = (jclass)e->NewWeakGlobalRef(e->FindClass("java/util/BitSet"));
    jmB = e->GetMethodID(jcB, "<init>","()V");
    jcI = (jclass)e->NewWeakGlobalRef(e->FindClass("java/lang/Integer"));
    jmV = e->GetStaticMethodID(jcI, "valueOf",
        "(Ljava/lang/String;)Ljava/lang/Integer;");
    jmS = e->GetMethodID(jcB, "set", "(I)V");
    jfV = e->GetFieldID(jcI, "value", "I");
    jcS = (jclass)e->NewWeakGlobalRef(e->FindClass("java/lang/System"));
    jcP = (jclass)e->NewWeakGlobalRef(e->FindClass
        ("java/io/PrintStream"));
    jmP = e->GetMethodID(jcP, "println", "(Ljava/lang/Object;)V");
    jfO = e->GetStaticFieldID(jcS, "out", "Ljava/io/PrintStream;");
    jout = e->NewWeakGlobalRef(e->GetStaticObjectField(jcS, jfO));
    }

JNIEXPORT void JNICALL Java_Bar_main
    (JNIEnv *e, jclass this_class, jobjectArray args){
    jobject jbs = e->NewObject(jcB,jmB);
    e->CallVoidMethod(jbs,jmS,e->GetIntField(
        e->CallStaticObjectMethod(jcI,jmV,
        e->GetObjectArrayElement(args,0)),jfV));
    e->CallVoidMethod(jout,jmP,jbs);
    }
```

**Figure 18.** Example Java Program Rewritten Cache Optimized in Java/C++/JNI

On the other end, use of operator++ with the Figure 3 version on class jtype objects prior to the two polymorphic Java/JNI method invokes resulted in a program execution of 140 RISC instructions. The best hand-optimized program execution time was then approximately 4% better than this operator++ try/catch Figure 3 program execution time.

The suggestive results above were re-verified on a PC platform with the OS vendor compiler's optimization settings. Here, the results ranged from C++WJ yielding 30% improvement over non-optimized JNI usage to C++WJ using `operator++` yielding approximately 3% additional cost over the best hand-optimized result.

The results in both cases suggest that C++WJ usage can add little to cpu execution costs versus hand optimized JNI usage. The extra costs are small with respect to an application's costs and with respect to costs of native code not written to maximally pre-cache JNI function results. The results also suggest that the C++WJ design decision to separate out jtypes from Jtypes (and jtype `operator++`) helps to narrow the performance gap.

# Related Work

The author knows of no other documented attempt at an improved C++ JNI.

Cygnus Native Interface (CNI)[13] is aimed at C++/Java integration. It is based on Java extensions to the GNU C++ implementation and is not based on JNI. CNI offers better efficiency than JNI at the expense of portability across JVMs. CNI offers better convenience than JNI with natural-to-C++ Java field access, method invoke, and object creation. In this latter respect, CNI is even more convenient to use than C++WJ. CNI, in its current stage, does not support some aspects of JNI including method call or field access through interface typed objects.

Beyond unimplemented features, CNI differs non-obviously with C++WJ in some aspects. For example, CNI uses C++ pointers to access Java objects. C++WJ uses C++ class objects here. (It is not obvious what implementation, if any, CNI will use to support Java downcasts from interfaces once interfaces are implemented.) Finally, CNI uses the C++ namespace feature to map Java package qualified class names. C++WJ uses JNI name encoding.

A separate effort aimed at creation of Java classes and `native` methods from C++ classes to perform *Java mirroring* has been used in an approach to ease the burden of reusing extensive C++ code.[14] Java mirroring starts from a C++ class to create a Java class with native methods corresponding to C++ members. Invoke of the Java native method causes first a lookup (or cache obtain) of a C++ object followed by C++ member function call or C++ data member access. There are issues with how to handle C++ multiple inheritance which Java does not support. In a sense, C++ mirroring is the opposite of C++WJ which generates a C++ class from a Java class. The two techniques can be combined. First Java mirroring is applied to create Java classes and then **jH** is applied to create C++WJ classes. The C++WJ classes make for easier JNI usage for the Java mirroring implementation code.

# Conclusion

Java is expected to co-exist with C/C++ for many years. JDK 1.1/1.2 JNI offered better portability than JDK 1.0 NMI but at the same time added complexity to native code JNI function calls. This paper has rationalized an approach to simplify and compile time type check 1.1/1.2 JNI usage by completely wrapping i.e. encapsulating JNI calls in C++. The result is a *step forward to the past* to bring back an easier JNI usage.

The C++ wrapped JNI technique presented is prototyped and is in use at Telcordia. Html reference documentation for the **jH** tool described in the paper is written. All mentioned caching techniques are employed.

Java and all Java-based trademarks are trademarks of Sun Microsystems, Inc.

# Acknowledgments

Steve Valin, Telcordia, provided suggestions on improving the technical presentation in this paper. George Phalen, Telcordia, provided assistance with non-technical issues.

# References

1. James Gosling, Bill Joy, Guy Steele, *The Java Language Specification*, Addison-Wesley, Reading, MA, 1996.
2. Bjarne Stroustrup, *The C++ Programming Language 2nd Edition*, Addison-Wesley, Reading, MA, 1991.
3. Dennis Ritchie, *The C Programming Language 2nd Edition*, Prentice-Hall, Upper Saddle River, NJ, 1989.
4. JTC 1/WG 21, ISO/IEC 14882:1998, Programming Languages -- C++.
5. JTC 1/WG 22, ISO/IEC 9899:1990, Programming languages -- C.
6. Sheng Liang, *The Java Native Interface*, Addison-Wesley, Reading, MA, 1999.
7. Sheng Liang, private communication.
8. Erich Gamma, Richard Helm, Ralph Johnson, John Vlissides, 'Design Patterns Elements of Resuable Object-Oriented Software', Addison-Wesley, Reading, MA, 1995.
9. A. Taivalsaari, 'On the notion of inheritance', ACM Computing Surveys, **28**(3), 438-479, (1996).
10. P. Wu, F. Wang, 'On Efficiency and Optimization of C++ Programs', Software Practice and Experience, **26**(4), 453-465, (1996).
11. Stanley B. Lippman, *Inside the C++ Object Model*, Addison-Wesley, Reading, MA, 1994.
12. D. Detlefs, A. Dosser, B. Zorn, 'Memory Allocation Costs in Large C and C++ Programs', Software Practice and Experience, **24**(6), 527-542, (1996).
13. Cygnus Solutions, *The Cygnus Native Interface for C++/Java Integration*, http://sourceware.cygnus.com/java/papers/cni/t1.html, 1999.
14. Rob Gordon, *Essential JNI*, Prentice Hall, Upper Saddle River, NJ, 1998.
15. Sheng Liang, Gilad Bracha, 'Dynamic Class Loading in the Java Virtual Machine', OOPSLA '98 Conference Proceedings, 36-44 (1998).
16. JavaSoft, Sun Microsystems, *Reflection*, http://java.sun.com/products/jdk/1.2/docs/guide/reflection, 1998.
17. G. Fowler, 'A case for make', Software Practice and Experience, **20(S1)**, S1/35-S1/46, (1990).